\documentclass[lettersize,journal]{IEEEtran}
\usepackage{amsmath,amsfonts}
\usepackage{algorithmic}
\usepackage{algorithm}
\usepackage{array}
\usepackage[caption=false,font=footnotesize,labelfont=rm,textfont=rm]{subfig}
\usepackage{textcomp}
\usepackage{stfloats}
\usepackage{url}
\usepackage{verbatim}
\usepackage{graphicx}
\usepackage{cite}
\usepackage{bm}
\usepackage{setspace}
\usepackage{multirow}
\usepackage{color}
\begin{document}

\title{Vision Aided Environment Semantics Extraction and Its Application in mmWave Beam Selection}

\author{Feiyang~Wen, 
        Weihua~Xu, 
        Feifei~Gao,
        Chengkang~Pan, 
        and Guangyi~Liu
\thanks{F. Wen, W. Xu, and F. Gao are with Institute for Artificial Intelligence, Tsinghua University (THUAI), Beijing National Research Center for Information Science and Technology (BNRist), Department of Automation, Tsinghua University, Beijing, P.R. China, 100084 
(email: wenfy20@mails.tsinghua.edu.cn, 
xwh19@mails.tsinghua.edu.cn, 
feifeigao@ieee.org).}
\thanks{C. Pan and G. Liu are with China Mobile Research Institute, Beijing, China. (e-mail: panchengkang@chinamobile.com; liuguangyi@chinamobile.com).}
}

\markboth{}
{Shell \MakeLowercase{\textit{et al.}}: A Sample Article Using IEEEtran.cls for IEEE Journals}

\IEEEpubid{}

\maketitle

\begin{abstract}
In this letter, we propose a novel mmWave beam selection method based on the environment semantics extracted from user-side camera images. 
Specifically, we first define the environment semantics as the spatial distribution of the scatterers that affect the wireless propagation channels and utilize the keypoint detection technique to extract them from the input images.
Then, we design a deep neural network with the environment semantics as the input that can output {the optimal beam pairs at the mobile station (MS)} and the base station (BS). 
Compared with the existing beam selection approaches that directly use images as the input, the proposed semantic-based method can explicitly obtain the environmental features that account for the propagation of wireless signals, thus reducing the storage and computational burden. 
Simulation results show that the proposed method can precisely estimate the location of the scatterers and outperform the existing works based on computer vision or light detection and ranging (LIDAR).
\end{abstract}


\section{Introduction}
\IEEEPARstart{M}{ILLIMETER} wave (mmWave) with beamforming is a critical technology for next-generation wireless communications and can achieve a higher transmission rate.
However, the traditional beamforming approaches, e.g., pilot-based strategies \cite{estimation} or beam sweeping \cite{sweeping}, are bottlenecked by a high spectrum overhead,
especially in time-varying scenarios such as vehicle-to-infrastructure (V2I) communications.
Recently, it has been indicated that with the aid of environment sensing information, such as point clouds \cite{lidar}-\cite{lidarpos_baseline} and images \cite{visionpos}-\cite{visionblockage}, one can implement beamforming with lower latency and fewer spectrum resources.

In \cite{lidar}, the authors utilize a deep neural network (DNN) classifier to predict the optimal beam pairs from the point clouds scanned by light detection and ranging (LIDAR).
The authors of \cite{lidarfed} and \cite{lidarpos_baseline} optimize the LIDAR-based method with federated learning strategies and various deep learning techniques.
With the aid of images taken at the base station (BS) side, \cite{visionpos} presents a multi-modal beam prediction scheme, and \cite{visionblockage} proposes a strategy for proactive blockage prediction and user hand-off.
Based on images from the camera view of the mobile station (MS), the authors of \cite{visionccm} perform channel covariance matrix estimation, and the authors of \cite{vbala} propose a beam alignment method supported by the object detection techniques.

However, the existing vision-based approaches are implemented in a straightforward manner, i.e., the optimal beam pairs are predicted directly from the input images, which may lead to certain disadvantages.
According to the discrete physical model for wireless propagation \cite{channelmodel}, among all the environmental characteristics, only a few scatterers can significantly affect the wireless channel.
Hence, there is a large amount of redundancy in the images, which reduces the training efficiency, causes the overfitting problem, and jeopardizes the accuracy of the beam selection.
Moreover, without explicitly extracting the effective scatterers, the existing methods have to use more complicated DNN models, resulting in a larger computational and storage burden.
To accurately represent the propagation environment, the semantic information of the images, i.e., the characteristics that account for the propagation of the wireless signal, should be extracted. Therefore, we define \textit{environment semantics} by considering the spatial distribution of the effective scatterers.

In this letter, we propose a vision-aided environment semantics extraction method and apply it to beam selection for V2I communication scenarios. 
In order to eliminate the redundant information, we represent the environment semantics as \textit{semantic heatmaps} and extract the heatmaps from the input images.
Then, the optimal beam pair is predicted by another neural network from the heatmaps. 
Simulation results indicate that the proposed method can accurately capture the environment semantics and outperform the existing vision-based or LIDAR-based methods.

\section{System Model}
\subsection{Channel Model}
Consider a downlink V2I communication system where a stationary BS serves an MS on a vehicle in the mmWave band. 
Both BS and MS are equipped with a uniform planar array (UPA). The numbers of antennas are $N_B = N_{B}^a \times N_{B}^b$ and $N_M = N_{M}^a \times N_{M}^b$, respectively. 
The channel matrix $\mathbf{H}$ can be {represented} by the widely-adopted geometric channel model:
\begin{equation}
  \label{channelmatrix}
  \mathbf{H} = \sum_{p=1}^{P} \alpha_p \mathbf{a}_r(\theta_p^{A}, \varphi_p^{A}) \mathbf{a}_t^{\mathrm{H}}(\theta_p^{D}, \varphi_p^{D}),
\end{equation}
where $P$ is the number of multipath components (MPC), $\alpha_p$ denotes the complex gain of the $p$th propagation path, and
$\theta_p^{A}, \varphi_p^{A}, \theta_p^{D}, \varphi_p^{D}$ are the elevation and azimuth of the angle of arrival and departure of the $p$th path, respectively.
Moreover, the steering vectors $\mathbf{a}_r(\theta_p^{A}, \varphi_p^{A}) \in \mathbb{C}^{N_M \times 1}, \mathbf{a}_t(\theta_p^{D}, \varphi_p^{D}) \in \mathbb{C}^{N_B \times 1}$ are defined as
\begin{equation}
  \label{steeringvector}
  \begin{aligned}
  \mathbf{a}(N^a, N^b; \theta, \varphi) =& \frac{1}{\sqrt{N^aN^b}}[1, ..., e^{j\pi(a\cos{\theta} + b\sin{\theta}\sin{\varphi})},\\
   &..., e^{j\pi((N^a - 1)\cos{\theta} + (N^b - 1)\sin{\theta}\sin{\varphi})}].\\
  \end{aligned}
\end{equation}
Then, we have $\mathbf{a}_r = \mathbf{a}(N_M^a, N_M^b)$ and $ \mathbf{a}_t = \mathbf{a}(N_B^a, N_B^b)$.

Assume that BS and MS are equipped with fixed beam codebooks. The codebook at the transmitter is $\mathcal{W}_B = \{\mathbf{w}_B^1, \mathbf{w}_B^2, ..., \mathbf{w}_B^{C_B}\}$, while the codebook at the receiver is $\mathcal{W}_M = \{\mathbf{w}_M^1, \mathbf{w}_M^2, ..., \mathbf{w}_M^{C_M}\}$, where $\mathbf{w}_B^i \in \mathbb{C}^{N_B \times 1}, \mathbf{w}_M^j \in \mathbb{C}^{N_M \times 1}$ denote the transmitting and receiving vector.
Under the noise-free assumption, the received power gain of beam pair $(\mathbf{w}_B^i, \mathbf{w}_M^j)$ is
\begin{equation}
  \label{powergain}
  y_{ij} = |(\mathbf{w}_M^j)^{\mathrm{H}}\mathbf{H}\mathbf{w}_B^i|^2.
\end{equation}

The goal is to select the optimal beam pair from the {beam codebooks} that can maximize the power gain, i.e.,
\begin{equation}
  \label{selection}
  (\mathbf{w}_B^{i*}, \mathbf{w}_M^{j*}) = \mathop{\arg\max}_{\mathbf{w}_B^{i} \in \mathcal{W}_B, \mathbf{w}_M^{j} \in \mathcal{W}_M} y_{ij}.
\end{equation}

\subsection{Definition of the Semantic Heatmaps}
To represent the environment semantics, we define the effective scatterers and generate the semantic heatmaps by projecting the effective scatterers into the camera view of MS.

\begin{figure}[!t]
  \centering
  \subfloat[The definition of the effective scatterers.]{\includegraphics[width=3.0in, trim=0 0 0 0]{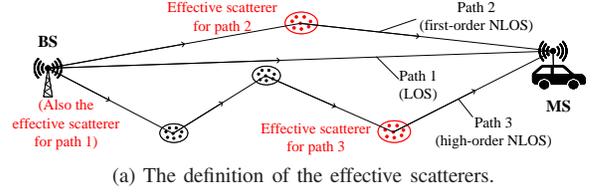}
  \label{fig_target}}
  \hfil
  \subfloat[The generation of the semantic heatmaps.]{\includegraphics[width=3.0in, trim=20 0 20 20]{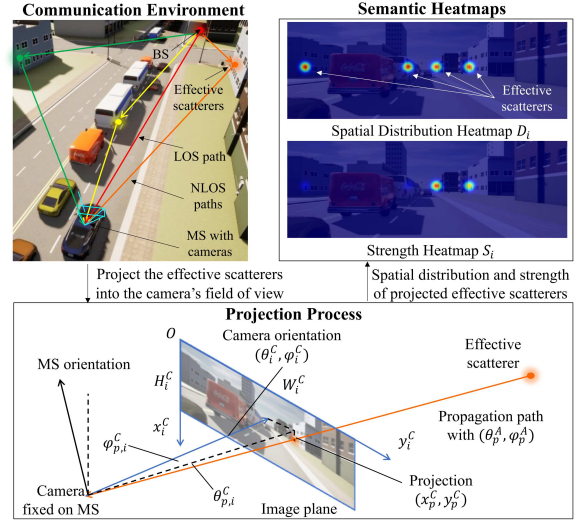}
  \label{fig_projection}}
  \caption{An illustration to the semantic heatmaps.}
  \label{fig_heatmap}
  \end{figure}

Note that the weak paths have less impact on the channel, and considering them would increase the computational and storage burden. 
Therefore, we neglect the paths with $|\alpha_p|^2/|\alpha_{max}|^2 < P_{th}$, where $|\alpha_{max}|^2$ denotes the power gain of the strongest path in a scene and $P_{th}$ is a tunable threshold to trade off the accuracy against the computational resources.
As shown in Fig. \ref{fig_heatmap}(a), for each none-line-of-sight (NLOS) path, we define the scatterer that reflects the last hop along the signal propagation as the \textit{effective scatterer};
while for the line-of-sight (LOS) path, BS serves as the effective scatterer.
In this way, we ensure that all the effective scatterers are in the camera view of MS and can be extracted from the images.

We assume that MS is equipped with $N_C$ monocular cameras, which are installed at different azimuths to provide multi-view images. 
Meanwhile, all the cameras are equipped at the same location with the antenna.\footnote[1]{In practice, if the cameras are fixed far away from the antenna, then one can utilize novel view synthesis techniques \cite{synthesis} to generate the antenna-view images from other camera views.} 
We define the elevation and the azimuth of the $i$th camera relative to MS as $(\theta_{i}^C, \varphi_{i}^C)$,
the size of the taken image as $H^C \times W^C$, and the horizontal field of view of the camera as $2\beta_i$.
If the $p$th effective scatterer is in the $i$th camera's field of view, then it is projected onto the {coordinates} $(x_{p}^C, y_{p}^C)$ of the image plane as
\setlength{\arraycolsep}{3pt}
\begin{equation}
  \begin{bmatrix} x_{p}^C \\
    y_{p}^C \\
    1\end{bmatrix}
    = \begin{bmatrix} \frac{W^C}{2\tan\beta_i} & 0 & H^C/2 \\
      0 & \frac{W^C}{2\tan\beta_i} & W^C/2 \\
      0 & 0 & 1\end{bmatrix}
      \begin{bmatrix} \frac{\tan\theta_{p,i}^C}{\cos\varphi_{p,i}^C} \\
        \tan\varphi_{p,i}^C \\
        1\end{bmatrix},
\end{equation}
where $\theta_{p,i}^C = \theta_{p}^A - \theta_{i}^C$ and $\varphi_{p,i}^C = \varphi_{p}^A - \varphi_{i}^C$ denote the elevation and the azimuth of the $p$th path in the $i$th camera's field of view.

Based on the projection process, we generate two types of heatmaps for each camera view: the spatial distribution heatmap $D_i$ and the strength heatmap $S_i$. Both heatmaps coincide with the image plane at a lower resolution, i.e., $D_i, S_i \in [0, 1]^{\frac{H^C}{R}\times \frac{W^C}{R}}$, and the coordinates of the $p$th scatterer are denoted as $(x_{p}^H, y_{p}^H) = (\lfloor\frac{x_{p}^C}{R}\rfloor, \lfloor\frac{y_{p}^C}{R}\rfloor)$.
The corresponding elements of the heatmaps are set to $D_{i}(x_{p}^H, y_{p}^H) = 1$ and $S_{i}(x_{p}^H, y_{p}^H) = |\alpha_p|^2/|\alpha_{max}|^2$, respectively.
Then, as a measure to improve the prediction performance of DNN, we {spread} all the projected points on $D_i$ and $ S_i$ using a Gaussian kernel $K = \exp{(\frac{(x - x_{p,i}^H)^2 + (y - y_{p,i}^H)^2}{2\sigma^2})}$ \cite{centernet}, where $\sigma$ is an empirical parameter determined by the size of the heatmaps.

\section{Vision Based Environment Semantics Extraction and Beam Selection}

\begin{figure*}[!t]
  \centering
  \includegraphics[width=6.6in, trim=0 15 0 15]{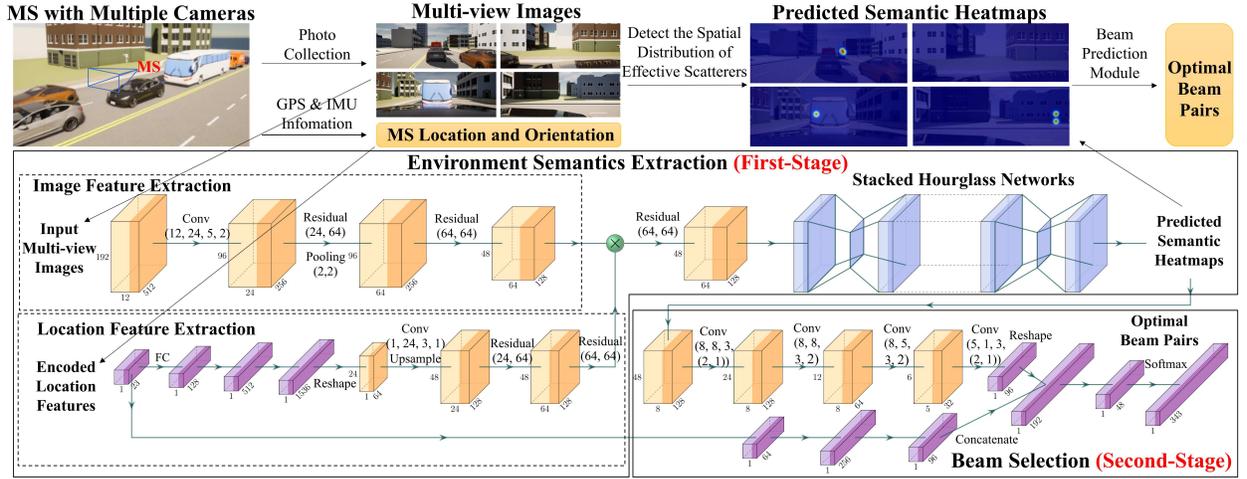}
  \caption{The proposed deep learning based method for environment semantics extraction and beam selection.}
  \label{fig_framework}
\end{figure*}

To obtain the optimal beam pairs in (\ref{selection}), 
we propose a two-stage beam selection approach, as shown in Fig. \ref{fig_framework}.
We assume that the location and the orientation of the MS can be obtained by vehicular sensors such as inertial measurement units and GPS receivers, and we assume that the location of the BS is known.
Therefore, the location $(x_M, y_M, z_M)$ and the orientation $\theta_M$ of the MS relative to the BS are available.

Different from \cite{visionpos}, \cite{visionblockage}, during the communication process, the optimal beam pairs are predicted by MS.
In the first stage, MS obtains its location information and multi-view images and sends them to a well-designed DNN for semantics extraction.
In the second stage, based on the semantic heatmaps and the location information, another DNN predicts the optimal beam pairs and feeds them back to the BS.
In the following {subsections}, we will present the two stages in detail.

\vspace{-2mm}
\subsection{Environment Semantics Extraction}
\subsubsection{Preprocessing}
At each time step, MS obtains the RGB images $I_1, I_2, ..., I_{N_C}$, where $I_i \in \mathbb{R}^{H^C\times W^C\times 3}$, and the location information that satisfies $(x_M, y_M, z_M, \theta_M) \in \mathbb{R}^4$.
We simply concatenate the images in the channel dimension to obtain the input image feature $I_{in}\in \mathbb{R}^{H^C\times W^C\times 3N_C}$.
However, as shown in \cite{nerf}, a DNN can poorly represent high-frequency features from the raw coordinate form of the location information, leading to limited performance. Hence, we encode $x_M, y_M$ using a high-frequency function \cite{nerf}:
\begin{equation}
  \gamma(p)\!\!=\!\!(\sin(2^0\pi p),\cos(2^0\pi p),...,\sin(2^{L-1}\pi p),\cos(2^{L-1}\pi p)),
\end{equation}
where $L$ is set to 5. The input location vector is concatenated as $L_{in} = (\gamma(x_M), \gamma(y_M), z_M, \cos\theta_M, \sin\theta_M) \in \mathbb{R}^{23}$.

\subsubsection{Semantics Extraction via Keypoint Detection}
The semantics extraction network takes $I_{in}, L_{in}$ as input and predicts the semantic heatmap $\hat{H} \in [0, 1]^{\frac{H^C}{R}\times\frac{W^C}{R}\times2N_C}$, where the first $N_C$ channels and the last $N_C$ channels correspond to $D_i$ and $S_i$ for each camera view, respectively.

As shown in Fig. \ref{fig_framework}, we use a convolutional layer and several residual blocks \cite{resnet} to extract the image feature map from $I_{in}$. Meanwhile, we use fully connected layers, 2-D reshape layer, and residual blocks to obtain the location feature map from $L_{in}$.
The image feature map can represent the interplay of environment scatterers, especially for dynamic scatterers like vehicles, while the location feature map contains the exact geometric relation between BS and MS.
We fuse the two feature maps using the multiplication operation to take full advantage of the contained information. 
Next, we adopt the stacked hourglass networks \cite{hourglass}, \cite{hourglass2} to predict $\hat{H}$. Generally, the hourglass networks capture multi-scale spatial relationships through a repeated compressing-upsampling architecture.

Denote $H$ as the groundtruth of $\hat{H}$ \footnote{During the training phase, the angle of arrival (AOA) and the power gains of the main propagation paths are required to generate $H$. In practice, they can be obtained via channel estimation or AOA estimation algorithms.}.
We use the {mean squared error (MSE)} loss $L_S= \sum_{x,y;C=N_C+1}^{2N_C} (\hat{H}_{xyC} - H_{xyC})^2$ for $S_i$ and a variation of focal loss \cite{cornernet}, \cite{centernet}
\begin{equation}
  L_D\!=\!-\!\!\!\!\!\sum_{x,y;C=1}^{N_C} \!\!\!
  \left\{
  \begin{aligned}
    &(1-\hat{H}_{xyC})^2\log(\hat{H}_{xyC}), \text{if} \ H_{xyC}=1\\
    &(1-H_{xyC})^4\hat{H}_{xyC}^2\log(1 - \hat{H}_{xyC}), \text{otherwise}
  \end{aligned}
  \right.
\end{equation}
for $D_i$, respectively. The heatmap prediction loss function is then defined as $L_{hm} = L_D + L_S$. 

\subsection{Beam Selection}
The beam selection network takes the predicted semantic heatmap $\hat{H}$ and the location vector $L_{in}$ as inputs.
We process $\hat{H}$ by several convolutional layers and process $L_{in}$ by fully connected layers. Next, two fully connected layers followed by a softmax layer are adopted to yield the prediction vector $\hat{y}_{ij} \in [0, 1]^{N_B\times N_M}$. The predicted optimal beam pair is then $(\hat{i}, \hat{j}) = \mathop{\arg\max}_{i, j} y_{ij}$.

The loss function for beam prediction is defined as \cite{lidarpos_baseline}
\begin{equation}
  L_{pd} = - (1 - \beta)\sum_{i,j} y_{ij}^*\log{(\hat{y}_{ij})} - \beta\sum_{i,j} \overline{y}_{ij}\log{(\hat{y}_{ij})},
  \label{eq_pdloss}
\end{equation}
where $\overline{y}_{ij}$ is the normalized form of the power gain $y_{ij}$; $y_{ij}^*=1$ for $(i^*, j^*)$ and $y_{ij}^*=0$ on other components; $\beta \in [0, 1]$ is a tunable hyperparameter.
The overall loss function is then defined as $L_{all} = L_{pd} + L_{hm}$.

\section{Simulation Results}
\subsection{Simulation Setup}
\begin{figure}[!t]
  \centering
  \includegraphics[width=2.8in, trim=0 0 0 0]{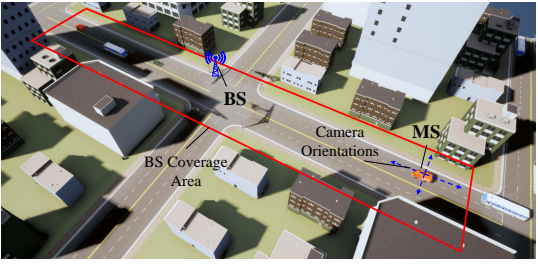}
  \caption{The simulation scenario in CARLA.}
  \label{fig_simulation}
\end{figure}

As shown in Fig. \ref{fig_simulation}, we delineate a rectangle with width $W_A$ and length $L_A$ as the coverage area of BS in the autonomous driving simulation software CARLA \cite{carla}.
Then, we utilize the traffic simulation software SUMO \cite{sumo} to generate a continuous traffic flow, which contains 5 types of vehicles as listed in TABLE \ref{tab_vehicle}.
The vehicles move according to the routes and velocities planned by SUMO, and the locations of the vehicles are synchronized to CARLA. 
Once a vehicle moves within the coverage area of BS, the cameras on the vehicle will take images at a regular time interval $T_C$, generating a sequence of images as the vehicle passes through the coverage area.
Meanwhile, the entire simulation scene is recorded and synchronized to the ray tracing software Wireless Insite \cite{wirelessinsite} for wireless channel simulation, in which all the vehicles are converted to cubes with the same sizes, locations, and orientations as those in CARLA.
In this way, we generate the dataset containing the sequences of MS locations, images, and corresponding channel information.
The dataset is made available to the public \cite{dataset}.

\begin{table}[!t]
  \small
  \caption{Simulated Vehicle Types.}
  \centering
  \begin{tabular}{|c||c|c|c|c|c|}
  \hline
  Type & Car \#1 & Car \#2 & Car \#3 & Van & Bus\\
  \hline
  Length/m & 4.81 & 4.90 & 4.15 & 5.20 & 11.08\\
  \hline
  Width/m & 2.17 & 2.06 & 2.00& 2.62 & 3.25\\
  \hline
  Height/m & 1.52 & 1.48 & 1.38 & 2.48 & 3.33\\
  \hline
  \end{tabular}
  \label{tab_vehicle}
  \end{table}

\begin{table}[!t]
\small
\caption{Ray Tracing Parameters for Wireless Insite.}
\centering
\begin{tabular}{|c||c|}
\hline
Parameter & Value\\
\hline
Carrier Frequency & {60 GHz}\\
\hline
Propagation Model & X3D\\
\hline
Building Material & Concrete\\
\hline
Vehicle Material & Metal\\
\hline
Maximum Number of Reflections & 6\\
\hline
Maximum Number of Diffractions & 1\\
\hline
Maximum Paths Per Receiver Point & 25\\
\hline
\end{tabular}
\label{tab_raytracing}
\end{table}

We set $W_A=$ 48 m, $L_A=$ 192 m, and $T_C=$ 0.5 s; the UPA of BS is fixed at 3 m above the street; the UPA of MS is fixed at 0.1 m above the roof center of the vehicle; the cameras are installed horizontally with $N_C=$ 4, $\varphi_i^C=$ 0$^{\circ}$, 90$^{\circ}$, 180$^{\circ}$, 270$^{\circ}$ respectively, $2\beta_i=$ 90$^{\circ}$, and $H_C\times W_C=$ 192$\times$512 in pixels.
For the semantic heatmaps, we set $\frac{H_C}{R}\times \frac{W_C}{R}=$ 48$\times$128, $P_{th}=$ -10 dB, and $\sigma=$ 1.5. The ray tracing parameters are shown in TABLE \ref{tab_raytracing};
$N_B^a = N_M^a =$ 8, $N_B^b = N_M^b =$ 64, $C_B = C_M =$ 64; 
$\mathbf{w}_B^i = \mathbf{a}_t(\theta_t, \frac{2i - C_B - 1}{2C_B}\pi)$, $i = 1,2,...,C_B$ and $ \mathbf{w}_M^j = \mathbf{a}_r(\theta_r, \frac{2j - C_M - 1}{2C_M}\pi)$, $i = 1,2,...,C_M$, where $\theta_t=92^{\circ}$ and $\theta_r=88^{\circ}$ according to the heights of BS and MS.
Similar to \cite{lidar}, we select 343 candidate beam pairs, which have become optimal more than three times in the whole dataset.
We then divide the dataset based on sequences, i.e., we obtain the training samples and the test samples from different sequences. The training set contains 801 sequences with 13163 LOS samples and 4759 NLOS samples, while the test set contains 492 sequences with 3725 LOS samples and 3245 NLOS samples.

The number of channels, layers, and stacks of the hourglass networks are 64, 4, and 2, respectively; $\beta$ is set to 0.8. 
We utilize batch normalization for each convolutional layer or residual block. The details are presented in Fig. \ref{fig_framework}.

\vspace{-4mm}
\subsection{Numerical Results}

\begin{table}[!t]
  \small
  \caption{{Performance Comparision Between Different Methods.}}
  \centering
  \begin{tabular}{|c||c|c|c|c|}
  \hline
  Model & {Proposed} & {LIDAR} & {VBALA} & {Location}\\
  \hline
  A(1) & {57.82\%} & {44.73\%} & {47.68\%} & {37.06\%}\\
  \hline
  T(1) & {80.14\%} & {67.69\%} & {72.16\%} & {57.47\%}\\
  \hline
  A(5) & {75.12\%} & {69.74\%} & {72.04\%} & {65.81\%}\\
  \hline
  T(5) & {90.14\%} & {86.94\%} & {88.61\%} & {83.97\%}\\
  \hline
  \end{tabular}
  \label{tab_performance}
  \end{table}

  \begin{figure}[!t]
    \centering
    \includegraphics[width=3.2in, trim=10 15 0 20]{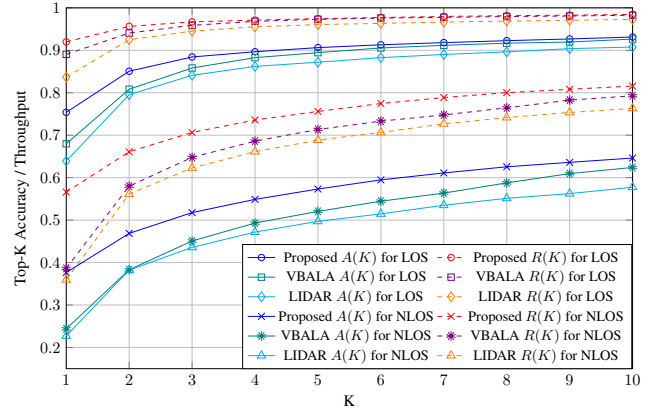}
    \caption{{Top-$K$ beam selection accuracy and throughput ratio of the three methods for LOS and NLOS test samples.}}
    \label{fig_result}
  \end{figure}

We compare the proposed method with the LIDAR-based method \cite{lidarpos_baseline} and the \textit{vision-based beam alignment when the MS location is available} (VBALA) \cite{vbala}. 
The LIDAR-based method performs beam selection from the point cloud and the MS location, 
while VBALA predicts the beam pairs from the MS location and the vehicle distribution feature captured by the object detection. 
For the LIDAR-based method, we equip MS with a LIDAR at 1 m above the roof center of the vehicle, which has a coverage radius of 100 m. For VBALA, we set the grid size $L_G=$ 12 m and $W_G=$ 2 m. 
Furthermore, we adopt a location-based beam selection method \cite{visionpos}, which only uses the location of the MS relative to the BS, as a baseline. For fairness, all the methods are trained for 30 epochs, using the same beam prediction loss function $L_{pd}$ in (\ref{eq_pdloss}).

Similar to \cite{lidarpos_baseline}, we evaluate the top-$K$ accuracy $A(K)$ with the throughput ratio $T(K)$. 
As shown in Fig. \ref{fig_result} and TABLE \ref{tab_performance}, the accuracy and {the} throughput ratio of the proposed method significantly outperform the baselines. 
Meanwhile, we compare the methods for LOS and NLOS test samples. Though the proposed method can outperform the baseline methods in LOS cases, its major advantage is exhibited in NLOS cases, where the top-1 accuracy is approximately 13.3\% higher than VBALA.
Since the environment semantics can be extracted in an explicit manner, the proposed method has a deeper perception of multipath effects and blockages, leading to a better NLOS performance.

\begin{figure}[!t]
  \centering
  \subfloat[The precision-recall curves. The distance threshold is 3 pixels, which corresponds to an angular misalignment of approximately 2.1$^{\circ}$.]{\includegraphics[width=2.9in, trim=20 0 0 30]{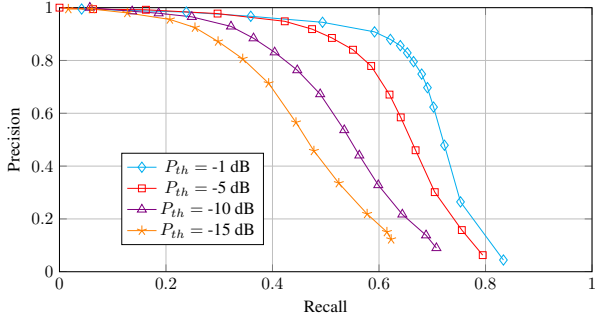}%
  \label{fig_precision_recall}}
  \hfil
  \subfloat[Visualization of spatial distribution heatmaps.]{\includegraphics[width=2.6in, trim=20 10 20 30]{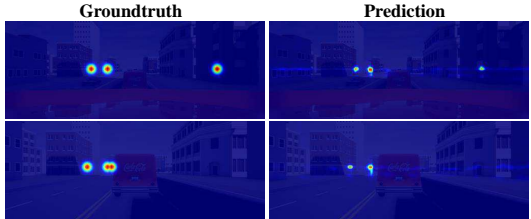}%
  \label{fig_distribution}}
  \hfil
  \subfloat[Visualization of strength heatmaps.]{\includegraphics[width=2.6in, trim=20 10 20 30]{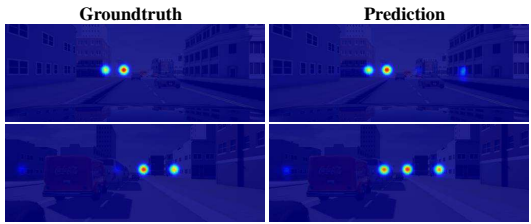}%
  \label{fig_strength}}
  \caption{The performance of semantics extraction. Assume the number of groundtruth, predicted, and successfully detected scatterers are $N_g, N_p, N_d$. The precision and the recall are defined as $N_d/N_p$ and $N_d/N_g$ respectively.}
  \label{fig_detection}
  \end{figure}

Moreover, we note that the threshold $P_{th}$ can affect the accuracy and the storage overhead of the proposed method. 
Specifically, we define the averaged number of effective scatterers per camera view as $N_E$, which reflects the storage requirement of the training dataset.
As shown in TABLE \ref{tab_threshold}, the proposed method achieves the highest accuracy at $P_{th} = -$10 {dB}. In addition, when $P_{th} < -$5 {dB}, the consumption of storage resources is significantly reduced without a major loss of accuracy.

\begin{table}[!t]
  \small
  \caption{{Performance Comparision Between Different Thresholds.}}
  \centering
  \begin{tabular}{|c||c|c|c|c|c|}
  \hline
  $P_{th}$ & {-1 dB} & {-5 dB} & {-10 dB} & {-15 dB}\\
  \hline
  A(1) & {56.84\%}  & {57.55\%} & {57.82\%} & {57.76\%} \\
  \hline
  T(1) & {79.27\%}  & {80.06\%} & {80.14\%} & {79.71\%} \\
  \hline
  $N_E$ & {0.328}  & {0.495} & {0.793} & {1.231} \\
  \hline
  \end{tabular}
  \label{tab_threshold}
  \end{table}

Furthermore, we evaluate the effectiveness of the environment semantics extraction in Fig. \ref{fig_detection}. We obtain the maximum points on the predicted spatial distribution heatmaps by non-maximum suppression. If the distance between the predicted maximum point and the groundtruth is within a threshold, then the corresponding scatterer will be perceived as detected.
Besides, since the stronger paths can more significantly influence the channel, the corresponding scatterers should be extracted more precisely. Therefore, we adjust the threshold $P_{th}$, train the corresponding semantics extraction model, and evaluate the precision-recall curves.
The proposed method is shown to successfully extract the effective scatterers from the input images, especially for the stronger ones.

\section{Conclusion}
We have proposed a vision-aided environment semantics extraction method and applied it to mmWave beam selection. We define the environment semantics as the spatial distribution of the effective scatterers, which are represented as the semantic heatmaps, and we extract them via keypoint detection techniques.
Compared with the existing vision-based or LIDAR-based methods, the proposed method shows deeper insight into the wireless propagation environment.
Simulation results show that the proposed method significantly outperforms the existing methods in accuracy and leads to lower overhead for beam selection.

\vfill
\end{document}